\documentclass[twoside,twocolumn,9pt]{article}
\usepackage{extsizes}
\usepackage[super,sort&compress,comma]{natbib} 
\usepackage[version=3]{mhchem}
\usepackage[left=1.5cm, right=1.5cm, top=1.785cm, bottom=2.0cm]{geometry}
\usepackage{balance}
\usepackage{times,mathptmx}
\usepackage{sectsty}
\usepackage{graphicx} 
\usepackage{lastpage}
\usepackage[format=plain,justification=justified,singlelinecheck=false,font={stretch=1.125,small,sf},labelfont=bf,labelsep=space]{caption}
\usepackage{float}
\usepackage{fancyhdr}
\usepackage{fnpos}
\usepackage[english]{babel}
\usepackage{array}
\usepackage{droidsans}
\usepackage{charter}
\usepackage[T1]{fontenc}
\usepackage[usenames,dvipsnames]{xcolor}
\usepackage{setspace}
\usepackage[compact]{titlesec}

\usepackage{epstopdf}

\definecolor{cream}{RGB}{222,217,201}

\begin{document}

\pagestyle{fancy}
\thispagestyle{plain}
\fancypagestyle{plain}{


\renewcommand{\headrulewidth}{0pt}
}

\makeFNbottom
\makeatletter
\renewcommand\LARGE{\@setfontsize\LARGE{15pt}{17}}
\renewcommand\Large{\@setfontsize\Large{12pt}{14}}
\renewcommand\large{\@setfontsize\large{10pt}{12}}
\renewcommand\footnotesize{\@setfontsize\footnotesize{7pt}{10}}
\makeatother

\renewcommand{\thefootnote}{\fnsymbol{footnote}}
\renewcommand\footnoterule{\vspace*{1pt}%
\color{cream}\hrule width 3.5in height 0.4pt \color{black}\vspace*{5pt}} 
\setcounter{secnumdepth}{5}

\makeatletter 
\renewcommand\@biblabel[1]{#1}            
\renewcommand\@makefntext[1]%
{\noindent\makebox[0pt][r]{\@thefnmark\,}#1}
\makeatother 
\renewcommand{\figurename}{\small{Fig.}~}
\sectionfont{\sffamily\Large}
\subsectionfont{\normalsize}
\subsubsectionfont{\bf}
\setstretch{1.125} 
\setlength{\skip\footins}{0.8cm}
\setlength{\footnotesep}{0.25cm}
\setlength{\jot}{10pt}
\titlespacing*{\section}{0pt}{4pt}{4pt}
\titlespacing*{\subsection}{0pt}{15pt}{1pt}

\fancyfoot{}
\fancyfoot[RO]{\footnotesize{\sffamily{1--\pageref{LastPage} ~\textbar  \hspace{2pt}\thepage}}}
\fancyfoot[LE]{\footnotesize{\sffamily{\thepage~\textbar\hspace{3.45cm} 1--\pageref{LastPage}}}}
\fancyhead{}
\renewcommand{\headrulewidth}{0pt} 
\renewcommand{\footrulewidth}{0pt}
\setlength{\arrayrulewidth}{1pt}
\setlength{\columnsep}{6.5mm}
\setlength\bibsep{1pt}

\makeatletter 
\newlength{\figrulesep} 
\setlength{\figrulesep}{0.5\textfloatsep} 

\newcommand{\topfigrule}{\vspace*{-1pt}%
\noindent{\color{cream}\rule[-\figrulesep]{\columnwidth}{1.5pt}} }

\newcommand{\botfigrule}{\vspace*{-2pt}%
\noindent{\color{cream}\rule[\figrulesep]{\columnwidth}{1.5pt}} }

\newcommand{\dblfigrule}{\vspace*{-1pt}%
\noindent{\color{cream}\rule[-\figrulesep]{\textwidth}{1.5pt}} }

\makeatother

\twocolumn[
  \begin{@twocolumnfalse}
\vspace{3cm}
\sffamily
\begin{tabular}{m{4.5cm} p{13.5cm} }
& \noindent\LARGE{\textbf{THz Time-Domain Spectroscopy of Mixed CO$_2$--CH$_3$OH Interstellar Ice Analogs}}\\
\vspace{0.3cm} & \vspace{0.3cm} \\

&\noindent\large{\textbf{Brett A. McGuire,$^{\ast}$\textit{$^{a,b\ddag}$} Sergio Ioppolo,\textit{$^{c,d}$} Marco A. Allodi,\textit{$^{e,b}$} and
Geoffrey A. Blake$^{\ast}$\textit{$^{b,d}$}}}\vspace{0.5cm}\\

 & \noindent\normalsize{The icy mantles of interstellar dust grains are the birthplaces of the primordial prebiotic molecular inventory that may eventually seed nascent solar systems and the planets and planetesimals that form therein.  Here, we present a study of two of the most abundant species in these ices after water: carbon dioxide (CO$_2$) and methanol (CH$_3$OH) using TeraHertz (THz) time-domain spectroscopy and mid-infrared spectroscopy.  We study pure and mixed-ices of these species, and demonstrate the power of the THz region of the spectrum to elucidate the long-range structure (i.e. crystalline versus amorphous) of the ice, the degree of segregation of these species within the ice, and the thermal history of the species within the ice.  Finally, we comment on the utility of the THz transitions arising from these ices for use in astronomical observations of interstellar ices.}

\end{tabular}

 \end{@twocolumnfalse} \vspace{0.6cm}

  ]

\renewcommand*\rmdefault{bch}\normalfont\upshape
\rmfamily
\section*{}
\vspace{-1cm}


\footnotetext{\textit{$^{a}$~Address, Address, Town, Country. Fax: XX XXXX XXXX; Tel: XX XXXX XXXX; E-mail: xxxx@aaa.bbb.ccc}}
\footnotetext{\textit{$^{b}$~Address, Address, Town, Country. }}

\footnotetext{\dag~Electronic Supplementary Information (ESI) available: [details of any supplementary information available should be included here]. See DOI: 10.1039/b000000x/}

\section{Introduction}

\footnotetext{\textit{$^{a}$~National Radio Astronomy Observatory, 520 Edgemont Rd, Charlottesville, VA USA 22903. E-mail: bmcguire@nrao.edu}}
\footnotetext{\textit{$^{b}$~Division of Chemistry and Chemical Engineering, California Institute of Technology, Pasadena, CA USA 91125. }}
\footnotetext{\textit{$^{c}$~Department of Physical Sciences, The Open University, Walton Hall, Milton Keynes MK7 6AA, UK }}
\footnotetext{\textit{$^{d}$~Division of Geological and Planetary Sciences, California Institute of Technology, Pasadena, CA USA 91125. E-mail: gab@gps.caltech.edu }}
\footnotetext{\textit{$^{e}$~Department of Chemistry, The Institute for Biophysical Dynamics, and the James Franck Institute, The University of Chicago, Chicago, IL USA 60637}}
\footnotetext{\ddag~B.A.M. is a Jansky Fellow of the National Radio Astronomy Observatory}

Cometary bombardment and meteoritic impacts have long been known to deliver substantial quantities of water and organic molecules to Earth, which may well have been the primordial prebiotic seeds of life.\cite{Chyba:1990yd}  This raises the question: \emph{What is the ultimate origin of this material?} While some chemical evolution can certainly occur \emph{in situ} in these icy bodies, a substantial portion of the molecular material is inherited directly from the parent molecular cloud.\cite{Cleeves:2014ty}  Thus, a thorough understanding of the primordial origins of our prebiotic molecular reservoir necessitates an examination of the genesis of this material in star- and planet-forming interstellar clouds.\cite{Belloche:2013eba}

Generally, simple, unsaturated molecules, as well as a number of long-chain hydrocarbons and fullerene species, can efficiently form \emph{via} gas-phase ion-molecule reactions\cite{Herbst:2009go}.  There is strong evidence, however, that more complex hydrogenated species, up to and including amino acids, are formed almost exclusively \emph{via} reactions on and within the icy surface of interstellar dust grains.\cite{Herbst:2009go,Garrod:2013id,Tielens:1982tb,Charnley:2001bt,Watanabe:2002od,Ioppolo:2008vi,Congiu:2012jw}  For example, the presence and abundance of methyl formate, one of the most prevalent interstellar complex organic molecules, has been argued to be explainable only through formation \emph{via} radical-radical recombination reactions in these icy bodies.\cite{Laas:2011yd}  Indeed, a recent laboratory study has shown that three abundant complex molecules -- methyl formate, glycolaldehyde, and ethylene glycol -- are efficiently formed in the solid phase through recombination of free radicals formed via H-atom addition and abstraction reactions that occur during the hydrogenation of CO ice at 15 K under dense molecular cloud conditions.\cite{Chuang:2015dl}

Despite their origins in molecular ices, the most complex molecule yet detected in the condensed-phase of the interstellar medium (ISM) is CH$_3$OH.\cite{Boogert:2015fx}  Indeed, only six species -- H$_2$O, CO, CO$_2$, CH$_3$OH, NH$_3$, and CH$_4$ -- have been securely identified observationally, although there is strong evidence for the additional presence of H$_2$CO, OCN$^-$, and OCS.\cite{Boogert:2015fx}  Thus, while characterizing these ices is critical for understanding the genesis of complex prebiotic material, we are currently limited in our ability to constrain models of chemical evolution in these condensed-phase environments where it occurs.

Much attention in the laboratory has been focused on the formation, destruction, and reaction of species within interstellar ice analogs, primarily using mid-infrared (mid-IR) spectroscopy.\cite{Linnartz:2015ec}  These studies, while crucial, have difficulty unambiguously measuring a critical component of the equation: the physical structure of the ice, which can have profound effects on reactions within the bulk material.\cite{Garrod:2013id}  Indeed, mid-IR spectroscopy is not the most powerful tool for examining this long-range structure, as, in general, the signals observed in the mid-IR only probe \emph{intra}-molecular modes which are characteristically perturbed by the surrounding ice structure.  In the far-IR, or TeraHertz (THz, 0.1 -- 10 THz, 30 - 3000 $\mu$m), region of the spectrum, however, it is the softest degrees of freedom of the ice (i.e. \emph{inter}-molecular modes) that are probed.\cite{Profeta:2011cz}  

These \emph{inter}-molecular modes offer a unique probe of ice structure (i.e. crystalline vs. amorphous ice). The thermal history of the ice is revealed as well, since the change in ice from amorphous to crystalline phases is a non-reversible process that starts at $\sim$110 K for water ice under laboratory (ultra) high vacuum conditions.\cite{Allodi:2013ma,Ioppolo:2014fd,Palumbo:1997cm,Mastrapa:2009hb,Jenniskens:1996wu,Moore:1992ys}  Recent observations of crystalline water ice have suggested this may be a powerful tool in studies of the evolution of planetary systems from the initial collapse phase through planet formation.\cite{McClure:2015kr}  The extreme sensitivity of the THz region to these structural modes opens the door to the study of species less-abundant than water, that are just as critical to our understanding of both physical and chemical evolution within forming systems.  The Far Infrared Field-Imaging Line Spectrometer (FIFI-LS) aboard the Stratospheric Observatory for Infrared Astronomy (SOFIA) offers bandwidth that is well-matched to these THz modes, covering 51 -- 203 $\mu$m (1.5 -- 5.9 THz) across two spectral bands.

The THz region of the spectrum has historically been challenging to access.  Recent advances in generation and detection techniques for THz photons, however, have allowed us to construct a broadband, sensitive, and coherent spectrometer whose spectral resolution is ideally-matched to the modes arising from the bulk motion of interstellar ice analogs.  We have previously reported on THz time-domain spectroscopy (THz-TDS) of pure, mixed, and layered ices of simple species (\ce{CO2}, \ce{H2O})\cite{Allodi:2013ma}, as well as more complex species (HCOOH, \ce{CH3COOH}, \ce{CH3CHO}, \ce{CH3OH}, and \ce{(CH3)2CO}).\cite{Ioppolo:2014fd}   Here, we present a comprehensive study of \ce{CO2}--\ce{CH3OH} mixtures in crystalline ices.  We examine the role of segregation within the ices on the spectra at various mixing ratios, and discuss the possible impacts on the utility of these spectra for comparisons to observations.

\section{Experimental Methods}

\begin{figure*}[h!]
\centering
\includegraphics[width=\textwidth]{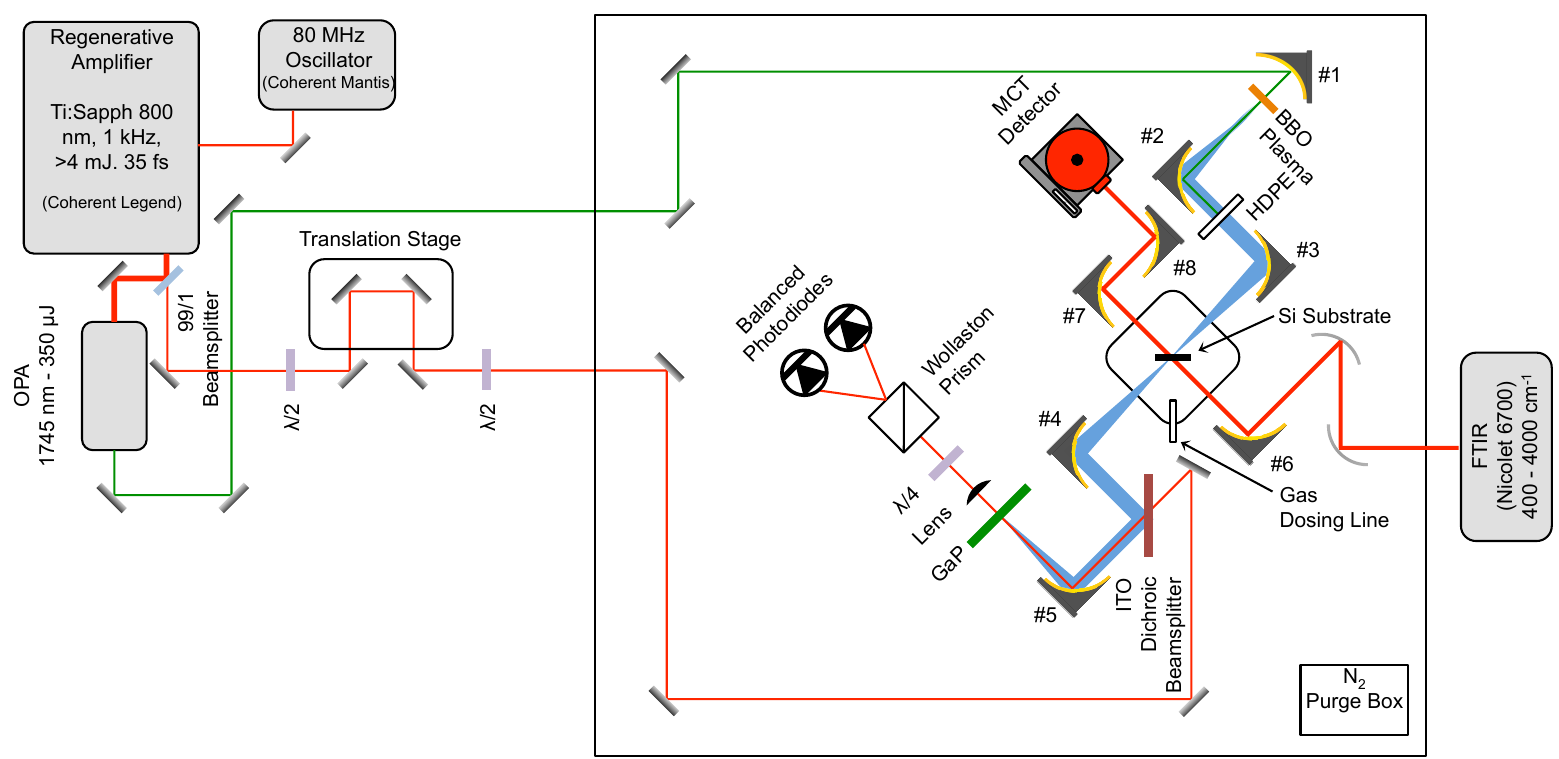}
\caption{Schematic overview of the Caltech THz-TD spectrometer, and its application to the study of astrochemical ice analogs. The 800 nm output of the Legend oscillator is split, with a portion of the light passing through an optical parametric amplifier (OPA).  This light is then doubled in a beta-barium borate (BBO) crystal and focused with an off-axis parabolic mirror (numbered optics) to spark a plasma.  A high-density polyethylene (HDPE) beam block filters out the visible light, and after passing through the sample, the THz beam is recombined with the other 800 nm beam in an indium tin oxide (ITO) dichroic beamsplitter and focused onto a gallium phosphide (GaP) crystal for detection.  The infrared spectrometer signal is detected by a mercury cadmium telluride (MCT) detector.}
\label{schematic}
\end{figure*}

The underlying principles of the experiment, as well as the technical details of the instrument, have been described in detail elsewhere,\cite{Allodi:2013ma,Ioppolo:2014fd}; a schematic is shown in Figure \ref{schematic}.  Briefly, a 35 fs, pulsed Ti:Sapphire regenerative amplifier at 800 nm drives an optical parametric amplifier (OPA) producing 1745 nm radiation in the idler beam.  A portion of this radiation is co-linearly doubled in a beta-barium borate (BBO) crystal, and the two pulses are focused in a dry N$_2$ purge, sparking a two-color plasma which produces intense, broadband THz radiation.\cite{Clerici:2013hx}  The THz light is then focused through the sample, recombined with a portion of the original 800 nm pulse in a gallium phosphide (GaP) crystal, and detected \emph{via} free-space electro-optic sampling.\cite{Wu:1995ec}  In this arrangement, the spectrometer provides coverage from $\sim$0.3 -- 7.0 THz.  Data were collected for 30 ps, producing an experimental resolution of $\sim$0.03 THz (1 cm$^{-1}$) when Fourier-transformed.  A commercially-available Fourier-transform infrared (FTIR) spectrometer provides simultaneous coverage from $\sim$500 -- 4000 cm$^{-1}$, also at 1 cm$^{-1}$ resolution.

To prepare the ices, gas-phase samples of CH$_3$OH and CO$_2$ were first mixed in the desired ratios in a 1 L glass bulb attached to the dosing line.  The pressures of each gas were monitored by a mass-independent pressure gauge.  Gas-phase CH$_3$OH was obtained by allowing a liquid sample of $\geq99.9\%$ CH$_3$OH (Sigma-Aldrich), which had been subjected to several freeze-pump-thaw cycles, to volatilize.  High-purity CO$_2$ from Air Liquide was used without further purification.  Once prepared and mixed, the samples were introduced into the chamber \emph{via} an all-metal leak valve, typically at a rate of $\sim$3.5 mTorr s$^{-1}$, to the desired total pressure ($P_{tot}$), where they were frozen onto a high-resistivity Si substrate held at $T_{dep}$ = 80 K.  In this high vacuum system, our ices are typically of order 10$^4$ monolayers (ML) thick.  After deposition, the samples were immediately cooled to a substrate temperature of $T_{sub}$ = 10 K, and spectra collected at 10 K, 20 K, and 30 K, followed by annealing, typically for $\sim$5 minutes, to $T_{ann}$ = 90 K, 120 K, and 140 K.  After each annealing, the samples were cooled to 10 K and spectra collected before the next annealing.  A detailed list of experiments is given in Table \ref{experiments}.

\begin{table}[h]
\small
\caption{Mixing ratios, total deposition pressure, deposition temperature, substrate temperatures, and annealing temperatures for ices described in this work.}
\label{experiments}
\begin{tabular*}{0.5\textwidth}{@{\extracolsep{\fill}}c c c c c}
\hline
CO$_2$:CH$_3$OH		&	$P_{tot}$ (Torr)		&	$T_{dep}$ (K)		&	$T_{sub}$	 (K)	&	$T_{ann}$ (K)	\\
\hline
1:3					&	5.7				&	80				&	10, 30, 60		&	90, 120, 140	\\
1:10					&	5.4				&	80				&	10, 30, 60		&	90, 120, 140	\\
\\
1:1					&	5.7				&	80				&	10, 30, 60		&	90, 120, 140	\\
\\
3:1					&	5.7				&	80				&	10, 30, 60		&	90, 120, 140	\\
10:1					&	5.7				&	80				&	10, 30, 60		&	90, 120, 140	\\
\\
CO$_2$				&	2.0				&	80				&	10, 30, 60		&	90, 120$^{\emph{a}}$, 140$^{\emph{a}}$	\\
CH$_3$OH			&	4.0				&	80				&	10, 30, 60		&	90, 120, 140	\\
\hline
\end{tabular*}
$^{\emph{a}}$After annealing to 120 K, the majority of the CO$_2$ ice had sublimed.  After the 140 K annealing, no CO$_2$ ice remained.
\end{table}

\section{Results}

\begin{figure}
\centering
\includegraphics[width=0.5\textwidth]{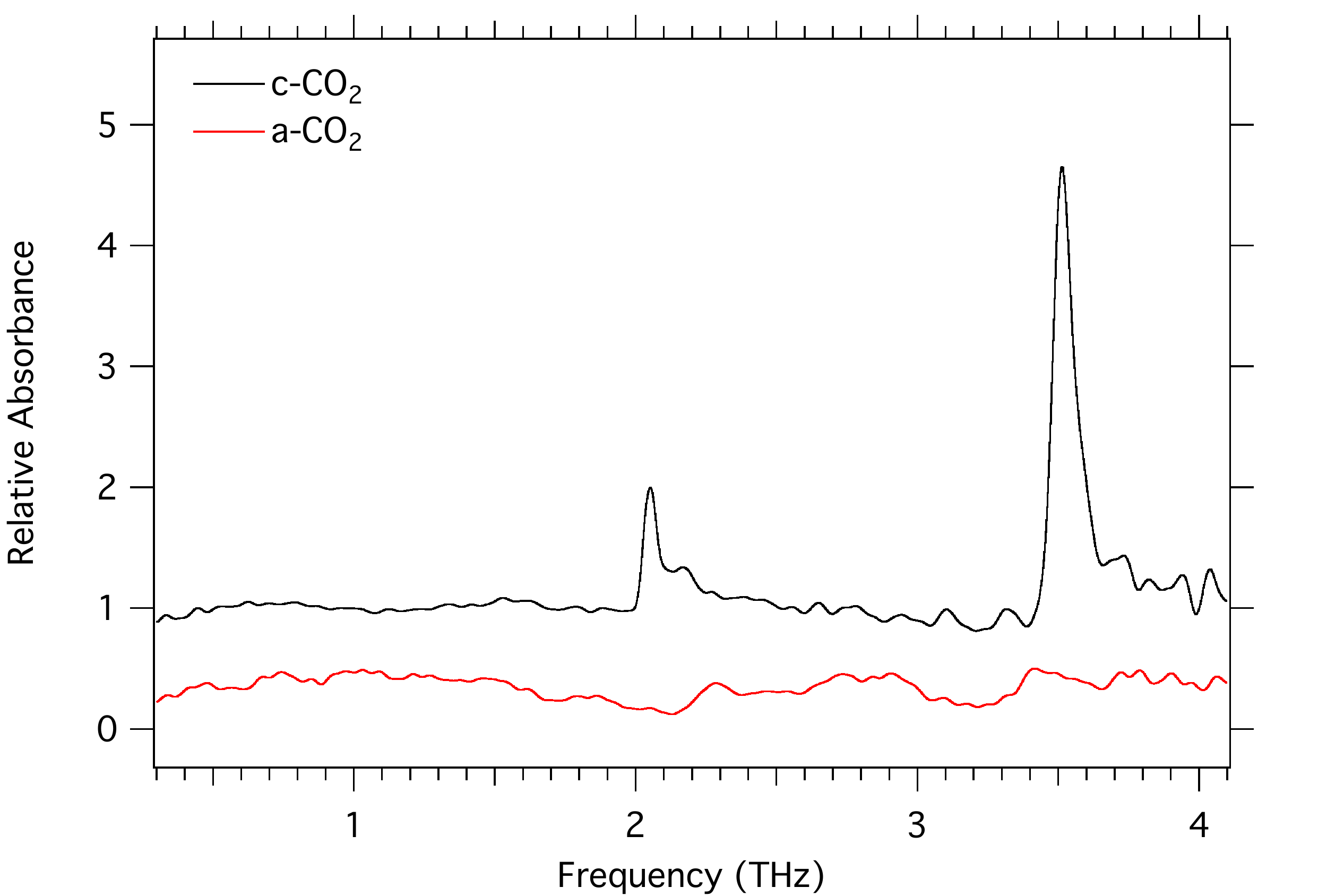}
\includegraphics[width=0.5\textwidth]{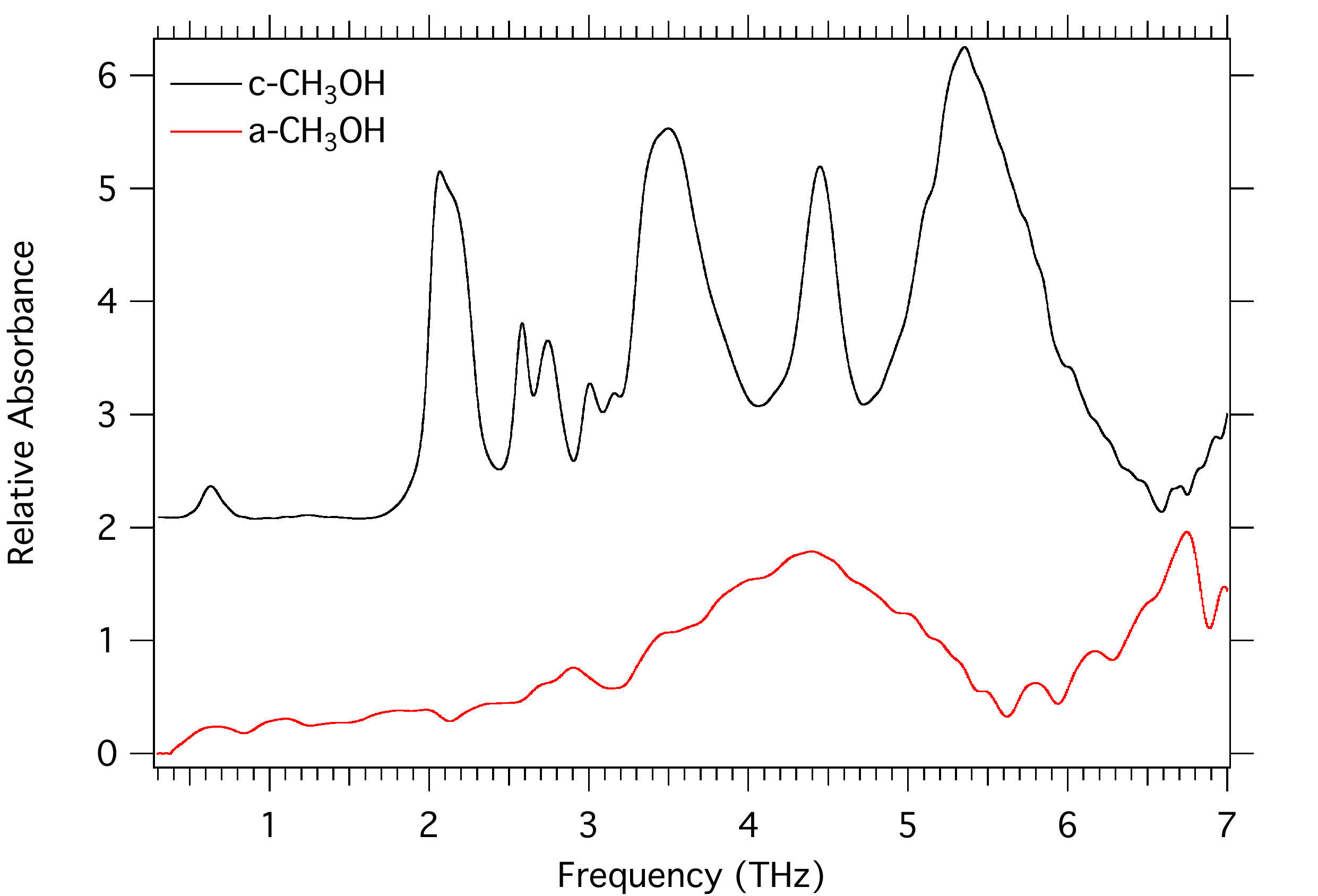}
\caption{(Top) spectra of amorphous CO$_2$ ($a$-CO$_2$) deposited at 10 K (red) and crystalline CO$_2$ ($c$-CO$_2$) deposited at 80 K (black).  Both spectra were acquired at 10 K, and have been vertically-offset for clarity.  (Bottom) spectra of $a$-CH$_3$OH deposited at 10 K (red) and $c$-CH$_3$OH deposited at 140 K (black).  Both spectra were taken at 10 K, and vertically-offset for clarity. }
\label{co2}
\end{figure}

We have previously reported on the temperature-dependent spectra of pure end-member crystalline methanol (herafter $c$-CH$_3$OH), \cite{Ioppolo:2014fd} but have re-measured the spectra for this study under identical temperature and annealing conditions for consistency.  Figure \ref{co2} shows that while $c$-CH$_3$OH ice is characterized by a series of sharper bands between 2 -- 6 THz, amorphous methanol ($a$-CH$_3$OH) is largely characterized by a broad feature around 4.3 THz, and the beginning of a second, broad signal around 6 THz.  Like the $a$-CH$_3$OH shown in Figure \ref{co2}, the $a$-CH$_3$OH ice generated in this study \emph{via} deposition at 80 K displays only a single, broad absorption across the 0.5 - 7 THz window.  An instrumental artifact around $\sim$2.1 THz is also seen in some scans.  The sharper, characteristic signals from $c$-CH$_3$OH become evident after annealing at 140 K, when sufficient energy is available to enable crystallization.  In a few cases, most prominently those where CH$_3$OH strongly dominates the CO$_2$, some $c$-CH$_3$OH features are seen at 120 K.  Notably, there does appear to be some weak signal from $c$-CH$_3$OH after the 120 K annealing step in the most dilute (10:1) CO$_2$:CH$_3$OH mixture.

Our first study of pure crystalline carbon dioxide ($c$-CO$_2$) was reported in \citet{Allodi:2013ma}, but a lack of sensitivity and resolution in these initial experiments showed no clear features, despite a reported prior observation of a feature at $\sim$3.3 THz in the literature.\cite{Moore:1994td}  More recently, \citet{Giuliano:2014ip} reported an observation of the 3.3 THz feature, as well as an additional feature around 2.1 THz.  However, they attribute these signals to amorphous carbon dioxide ($a$-CO$_2$) rather than $c$-CO$_2$.  We have recently conducted a thorough investigation of $a$-CO$_2$ and $c$-CO$_2$ features in this frequency range. In brief, we find unambiguous evidence that these two features are due solely to $c$-CO$_2$, and that $a$-CO$_2$ shows no distinct features within the coverage of our spectrometer.  For the purposes of this study, we have repeated the measurements of pure CO$_2$ under the same experimental conditions used for the mixtures.  For comparison, we also present a spectrum of $a$-CO$_2$ from the forthcoming Ioppolo \emph{et al.} publication (Fig. \ref{co2}).  We note that the sensitivity and resolution of the Caltech spectrometer has significantly improved in the two years since the publication of \citet{Allodi:2013ma} through a combination of instrumental upgrades.

\begin{figure*}
\centering
\includegraphics[width=0.33\textwidth]{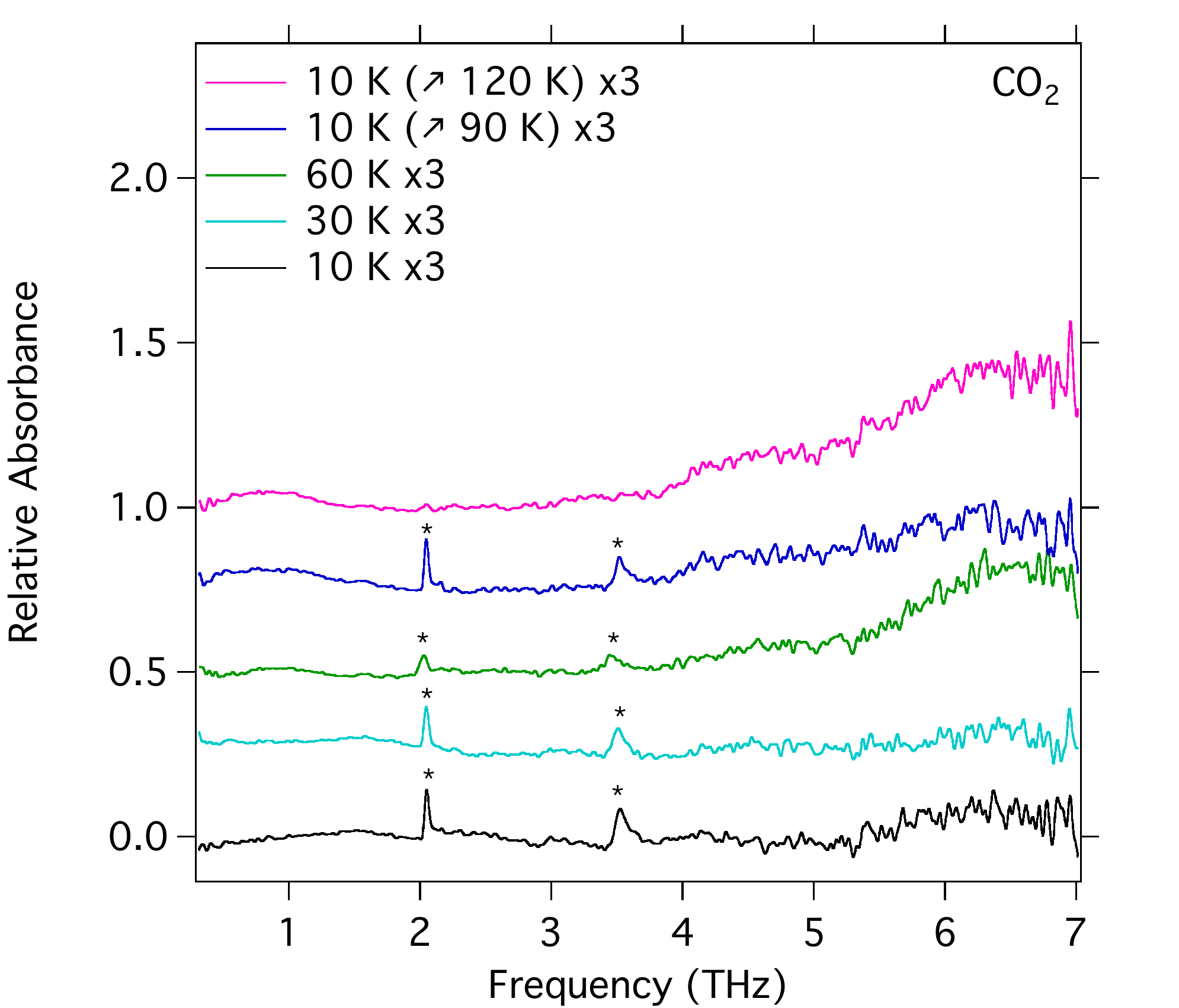}
\includegraphics[width=0.33\textwidth]{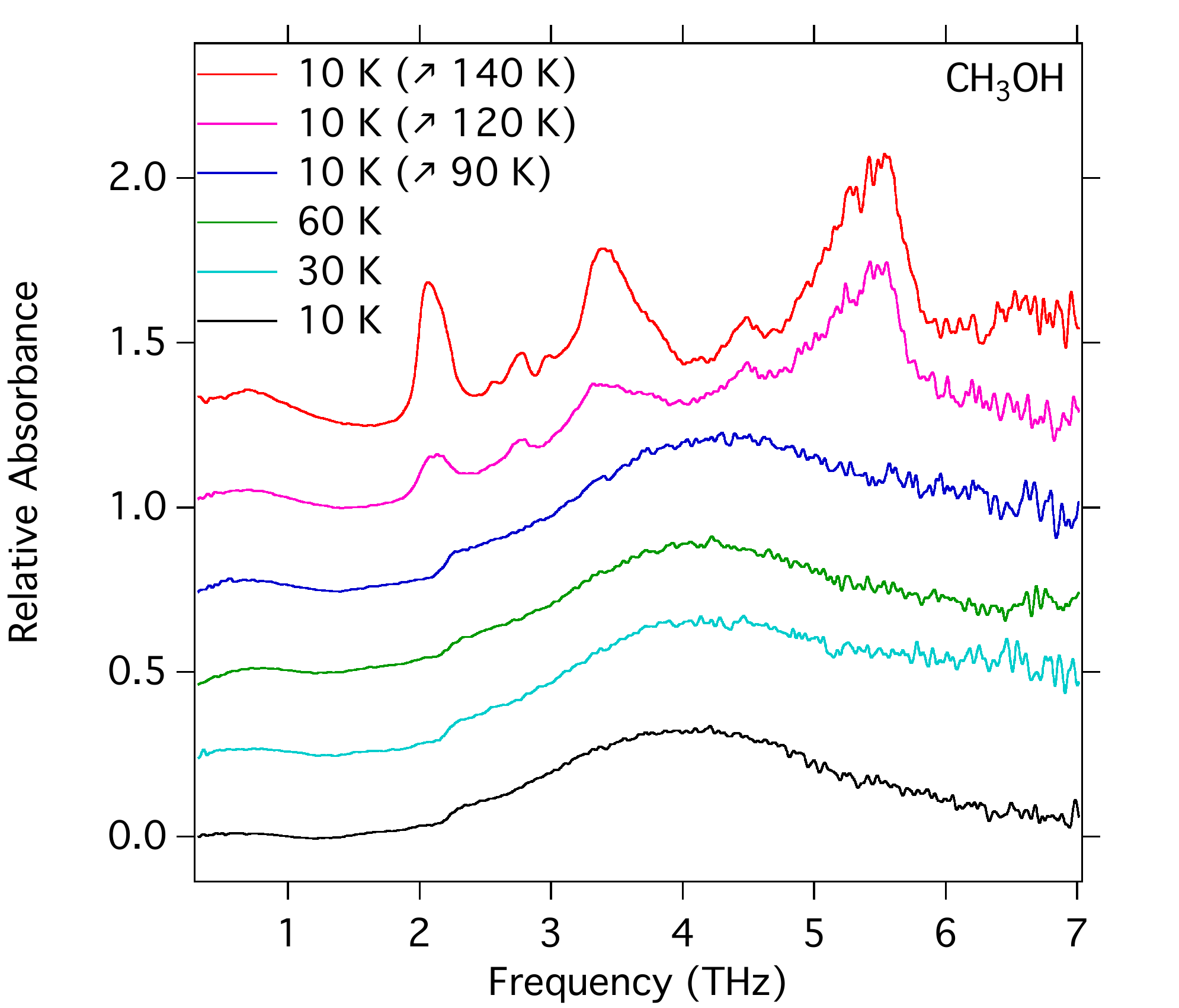}\\
\includegraphics[width=0.33\textwidth]{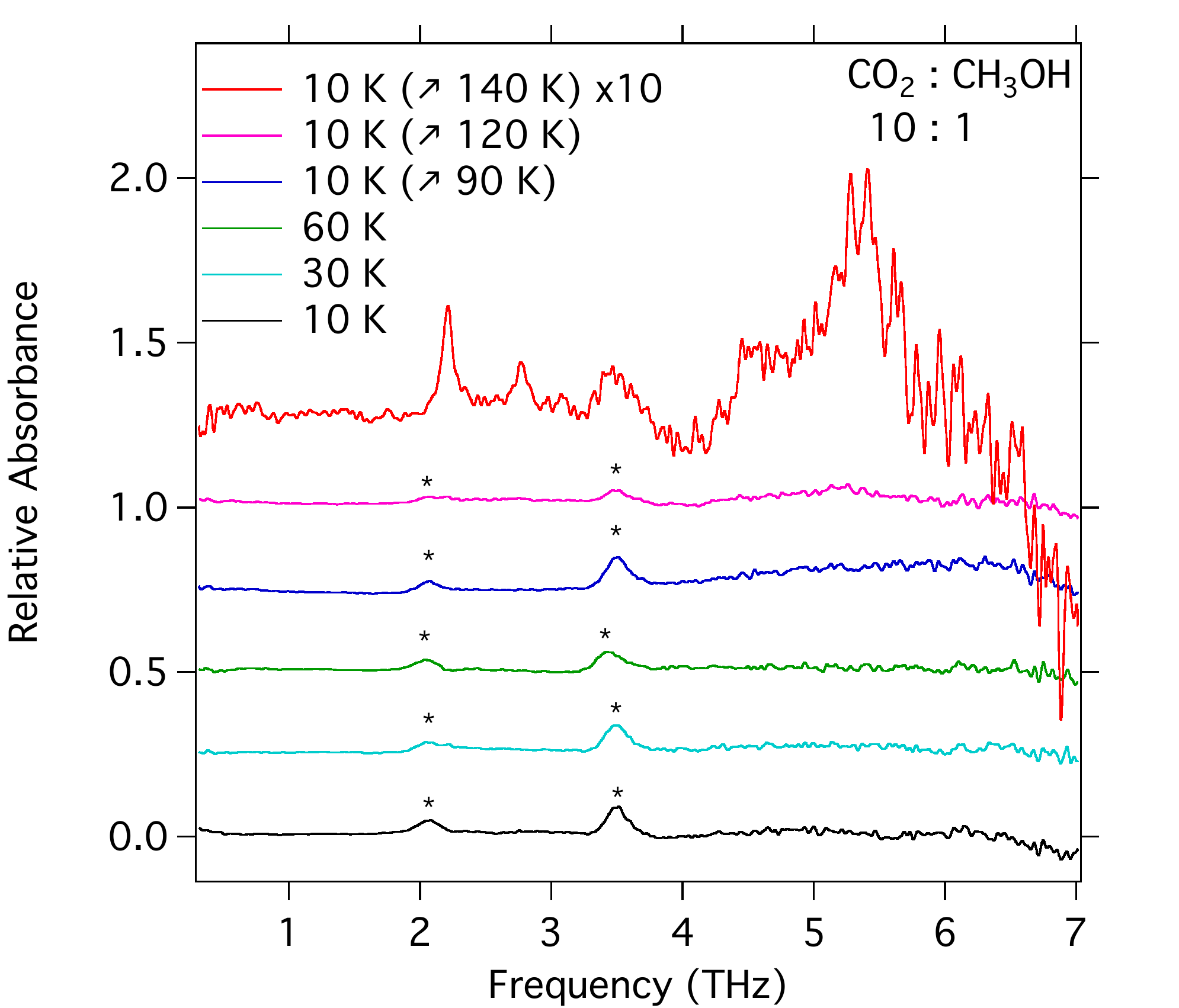}
\includegraphics[width=0.33\textwidth]{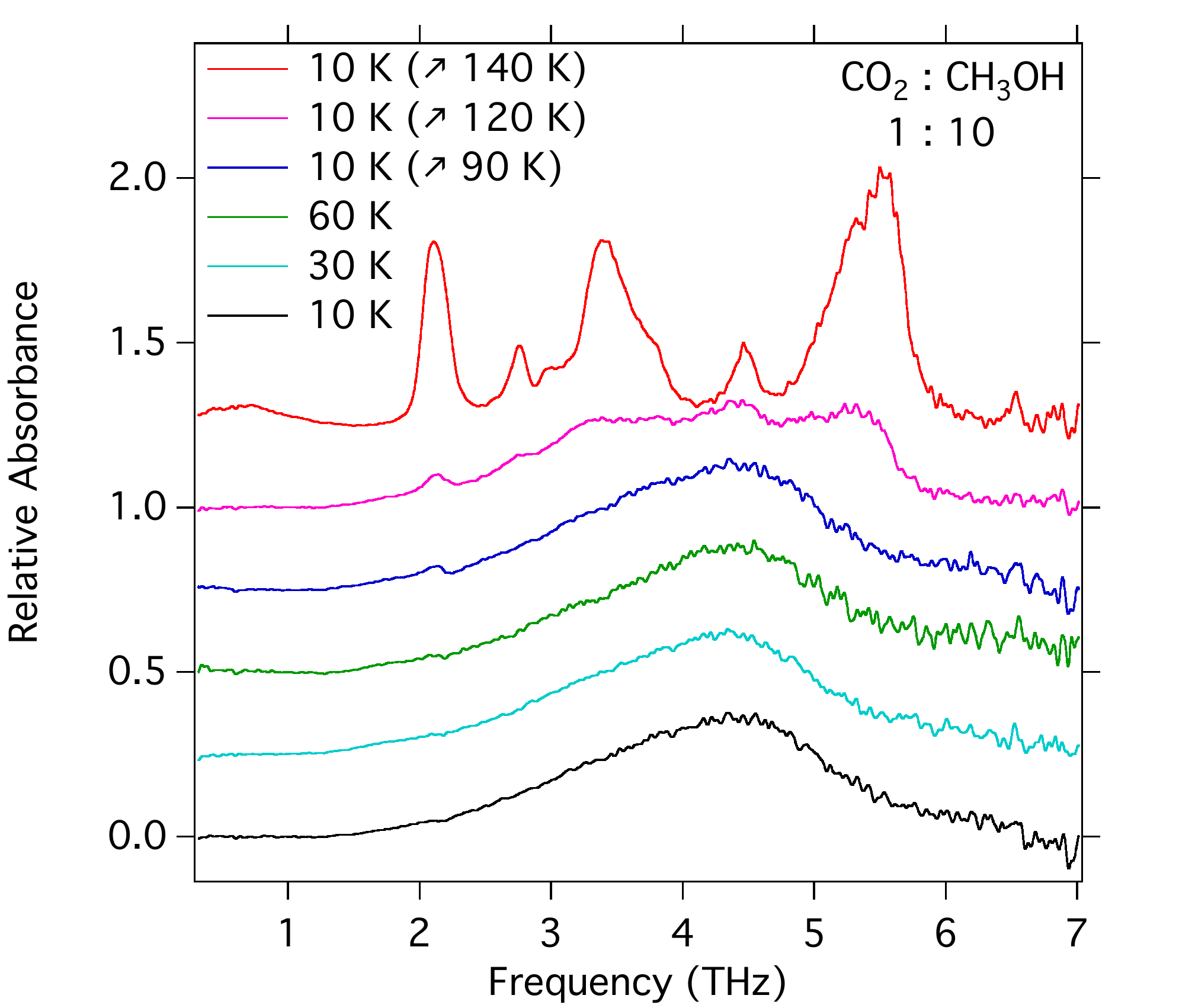}\\
\includegraphics[width=0.33\textwidth]{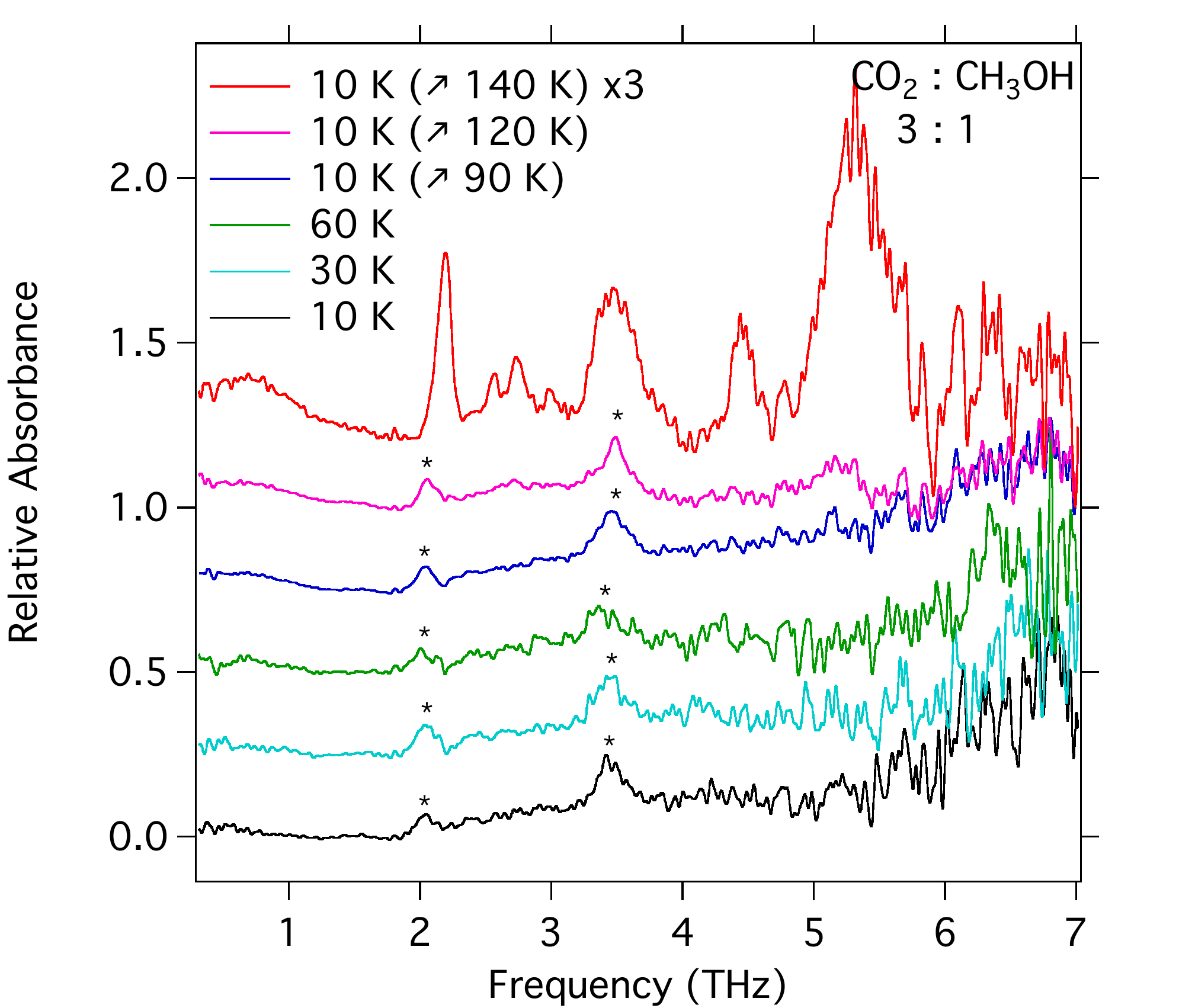}
\includegraphics[width=0.33\textwidth]{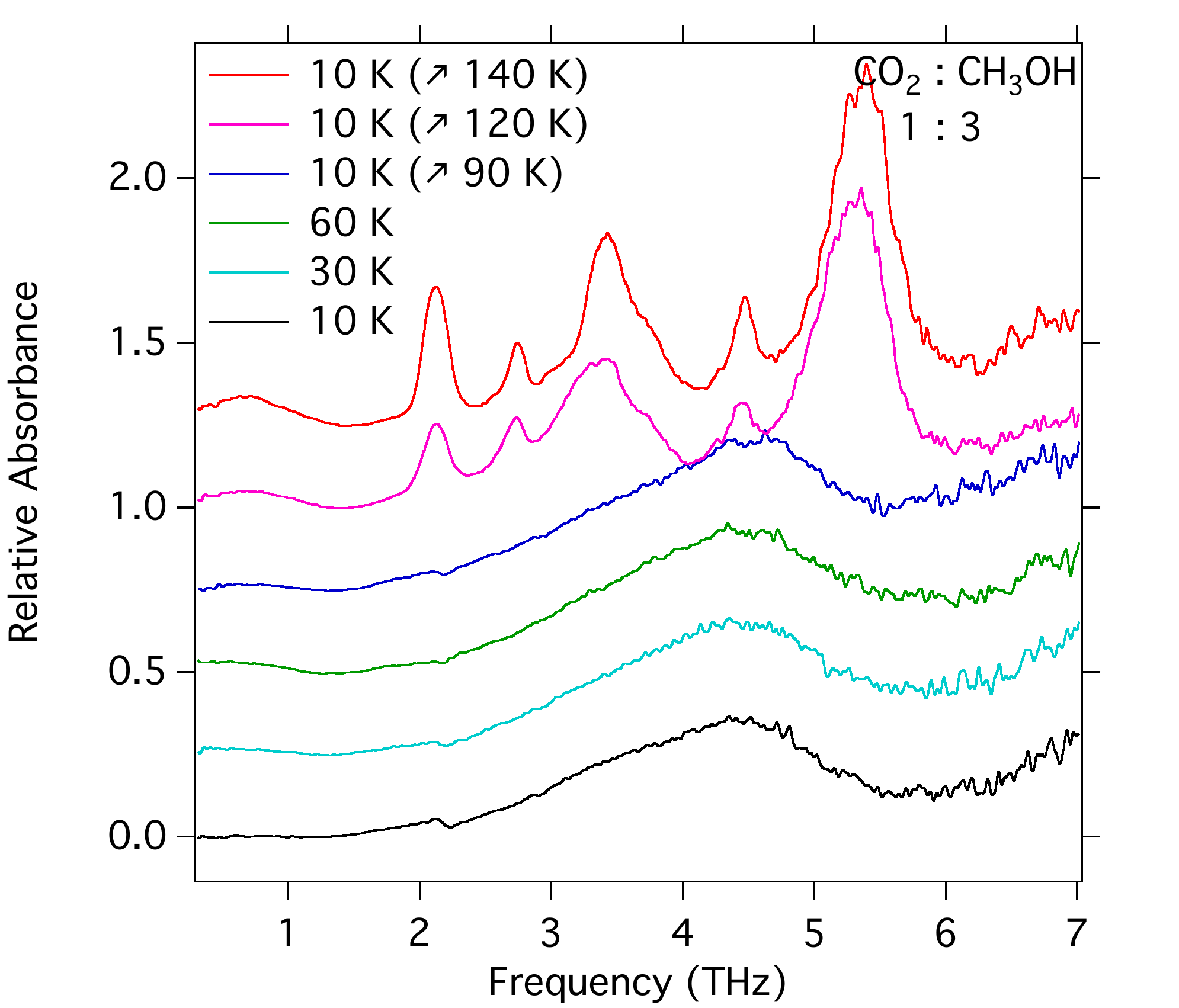}\\
\includegraphics[width=0.33\textwidth]{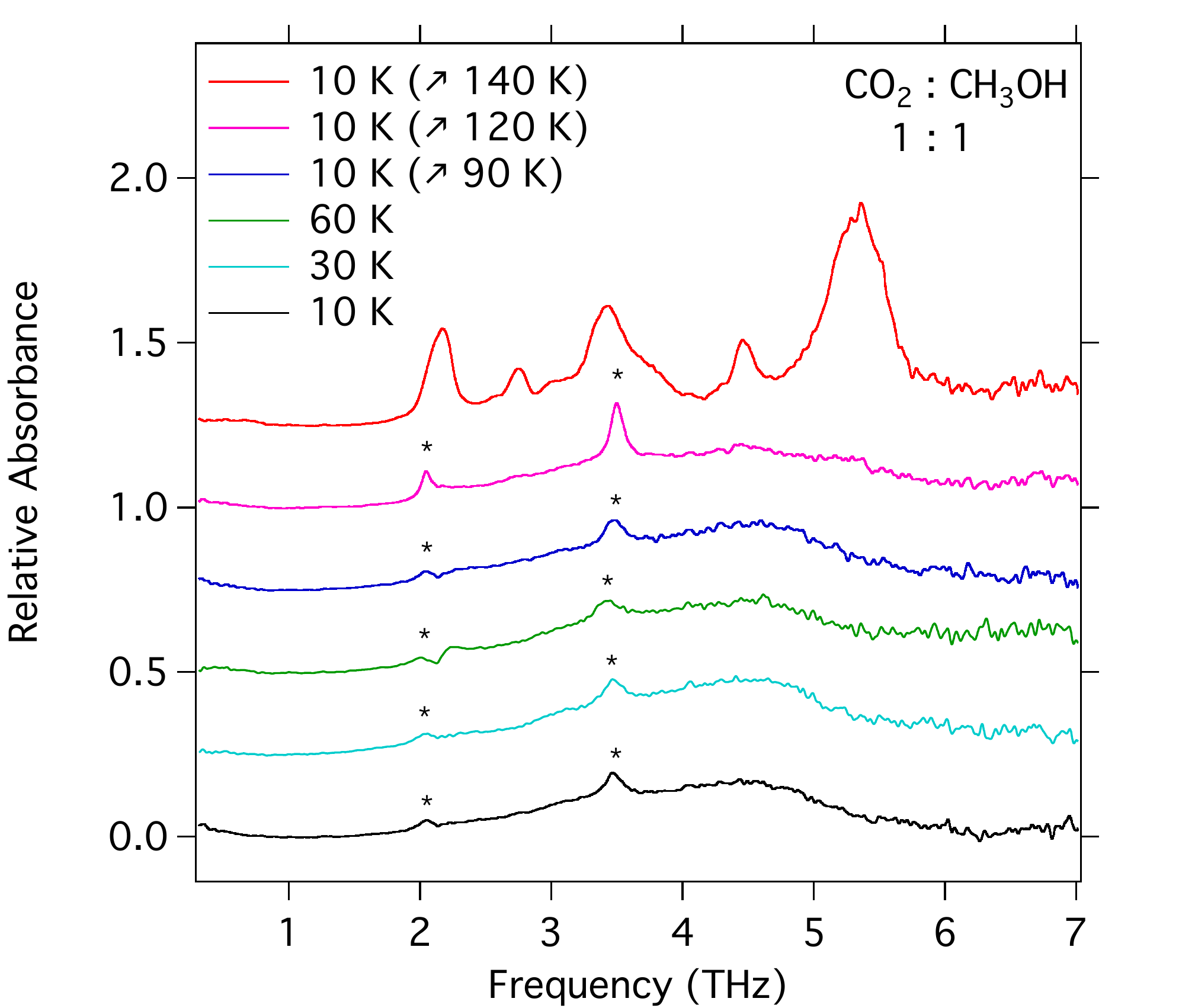}
\caption{THz-TDS spectra collected for this study.  All ices were deposited at 80 K, and spectra were collected at the temperatures indicated in the caption (10 K, 30 K, and 60 K).  In cases where the ices were annealed, the annealing temperature is indicated by $\nearrow$.  Ice compositions are given in the upper right of each panel.  Spectra are vertically-offset for clarity, and, when noted, are scaled to show detail.  The positions of the $c$-CO$_2$ features are marked by asterisks (*) when present.}
\label{spectra}
\end{figure*}

Spectra collected for this work are shown in Fig. \ref{spectra}.  The reduction procedure has been described in detail previously \cite{Ioppolo:2014fd}.  Briefly, a fast Fourier transform of the time-domain data is performed after applying an asymmetric Hann window to the data.  This converts the spectra to the frequency domain.  Baselines were then removed from the spectra by fitting the line-free regions to either a static offset or 1st-order linear fit, although the later was rarely required.

\subsection{CO$_2$-Dominated Mixtures}

Signals from the 2.1 THz and 3.5 THz $c$-CO$_2$ modes are clearly seen in the 1:1, 3:1, and 10:1 mixtures until annealing at 140 K.  In the case of pure $c$-CO$_2$, these features disappear after annealing at 120 K, unlike in the mixed cases.  Signal from $c$-CH$_3$OH is clearly seen after annealing to 140 K in all three mixtures, at which point no $c$-CO$_2$ is apparent, although the near-coincidence of the $c$-CO$_2$ features with two $c$-CH$_3$OH features makes this determination somewhat ambiguous.  For the 3:1 and 10:1 mixtures, features from $c$-CH$_3$OH, particularly at $\sim$5.2 THz and $\sim$2.6 THz, do begin to appear after annealing at 120 K.  The same features are possibly present in the 1:1 mixture, but would be just above the noise floor (baseline noise) if real.

\begin{figure}
\centering
\includegraphics[width=0.5\textwidth]{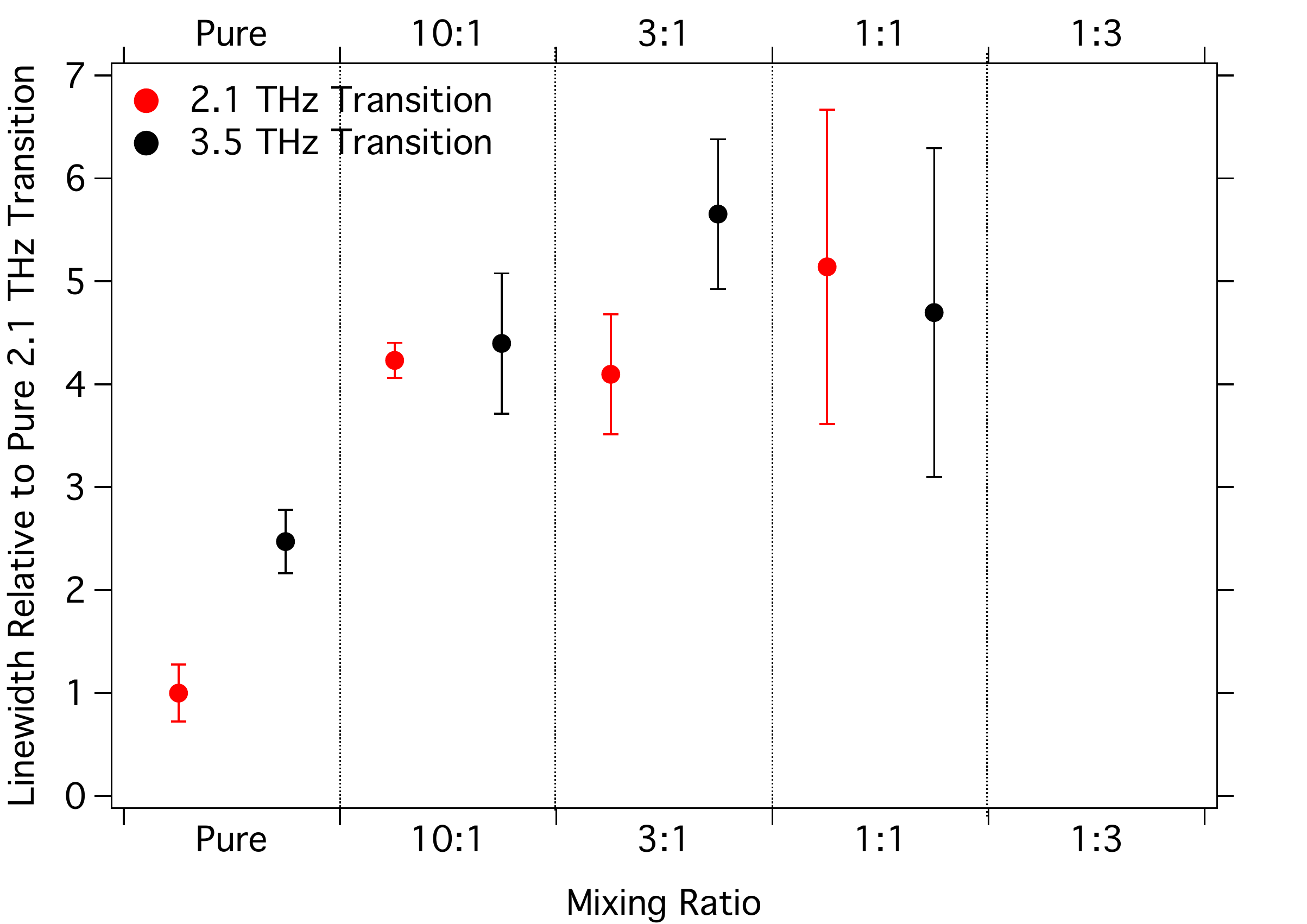}
\includegraphics[width=0.5\textwidth]{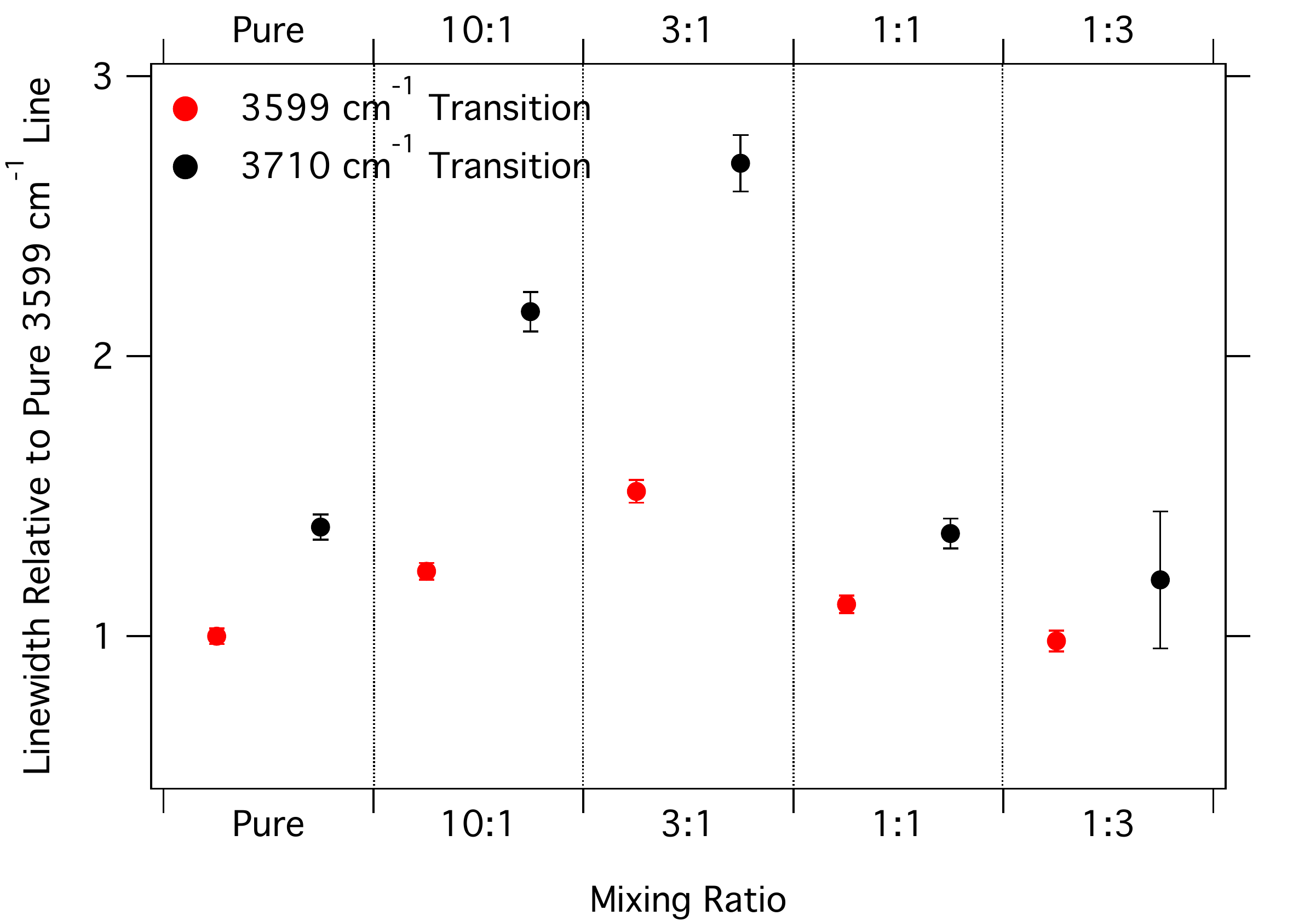}
\caption{(Top) Average FWHM values of the 2.1 THz and 3.5 THz $c$-CO$_2$ transitions observed in this work at 10 K, 30 K, and 60 K for pure CO$_2$ and the indicated mixing ratios.  Values are normalized to the pure, 2.1 THz width.  Error bars are 1$\sigma$ standard deviations in the averages. (Bottom) FWHM values of the 3599 cm$^{-1}$ and 3710 cm$^{-1}$ 2$\nu_2 + \nu_3$ and $\nu_1 + \nu_3$, respectively, CO$_2$ transitions at 10 K for pure CO$_2$ and the indicated mixing ratios.}
\label{co2widths}
\end{figure}

Figure \ref{co2widths} (top) shows a quantitative analysis of the linewidths of the observed $c$-CO$_2$ transitions, as determined by a Gaussian fit to the features. Due to the relatively low signal-to-noise ratio (SNR) of the lines, fits to the 10 K, 30 K, and 60 K signals were averaged to determine the linewidth, with the standard deviation of this average given as the error bars.  While some broadening of the transitions was observed with temperature, this broadening was not linear over the entire range of mixing ratios.  Instead, the linewidth appears to depend heavily on mixing ratio. The results show a significant increase in linewidth between pure CO$_2$ and mixed ices (Fig. \ref{co2widths}).  For the pure ice, the 2.1 THz transition is substantially narrower than the 3.5 THz transition.  The mixed ices, conversely, have a largely uniform linewidth regardless of mixing ratio (within the uncertainties).  They are also all significantly broader than either transition in pure CO$_2$.  The increase in linewidth between the pure and mixed 2.1 THz transitions, however, is markedly greater on average compared to the 3.5 THz transition: factors of 4.5 vs 2.0, respectively.

Figure \ref{co2widths} (bottom) shows a quantitative analysis of the linewidths of the observed CO$_2$ transitions in the FTIR shown in Figure \ref{ftirsets} (middle), as determined by a Guassian fit to the features.  Because of the much higher SNR, only the 10 K scan with no annealing was used for the determination, and the errors are purely due to uncertainty in the Gaussian fit.  As in the THz, the mid-IR linewidths show a mixing ratio dependence.  The magnitude of the change is about half that in the THz, and unlike the THz, seems to increase linearly for both transitions from pure CO$_2$ to the 3:1 mixture.  Finally, while the linewidths for the 1:1 mixture in the THz remain similar to those of the 10:1 and 3:1 mixtures, in the FTIR the 1:1 and 1:3 mixtures are significantly narrower.  

\subsection{CH$_3$OH-Dominated Mixtures}

The THz signatures of $c$-CO$_2$ appear to be strongly suppressed in both CH$_3$OH-dominated mixtures.  A very weak indication of the 2.1 THz $c$-CO$_2$ may be visible in a handful of scans, but it is difficult to distinguish from both the weak artifact (introduced by the HDPE beam block) in this region, and the overlapping $c$-CH$_3$OH mode which begins to appear after crystallization starts at 120 K.  Of the mixtures studied, only pure CH$_3$OH and the 3:1 CO$_2$ dominated mixture show a clear separation of the two $c$-CH$_3$OH transitions around 2.6 THz.  The others show only a single blended peak in this region. 

\section{Discussion}

After H$_2$O and CO, CO$_2$ is one of the most abundant ice species in the ISM, with abundances of nearly 20\% that of H$_2$O ice.\cite{Whittet:1998kd,Ehrenfreund:1999wp}  In dense molecular clouds, CO$_2$ is formed in the solid phase primarily through the CO + OH reaction that has been experimentally found to be 10 times more efficient than the CO + O channel.\cite{Ioppolo:2013bh}  Therefore, although CO is its parent molecule, CO$_2$ tends to reside primarily in polar ices (i.e. H$_2$O-rich rather than CO-rich), where OH radicals are more abundant and available for reaction with CO.  Thus, mid-IR and THz spectroscopic studies of CO$_2$ in polar environments are important to our understanding of the origin and evolution of interstellar CO$_2$ ices.\cite{Pontoppidan:2008ew,Allodi:2013ma}

Outside of mixtures with H$_2$O, CH$_3$OH is the most abundant polar ice constituent at $\sim$6--9\% of the abundance of H$_2$O, or $\sim$20--50\% of the abundance of CO$_2$.\cite{Boogert:2015fx}  CO and CO$_2$ have been shown to be the products of UV and cosmic ray irradiation of CH$_3$OH-containing ices.\cite{Islam:2014kc}  Therefore, in later stages of star formation, when ices are extensively exposed to heating, UV photons, and cosmic rays, CO$_2$ is thermally-processed and mixed with CH$_3$OH.\cite{Ioppolo:2013fg} As the segregation of CO$_2$ into ordered crystalline micro-domains is thought to be a powerful probe of thermal processing in astrophysical environments,\cite{Pontoppidan:2008ew,Kim:2012hz} it is interesting to explore that segregation in the laboratory, and its observational implications.

\subsection{Structure and Segregation}

The segregation of CO$_2$ ice upon deposition at warm temperatures has been previously reported in mixed CO$_2$--H$_2$O\cite{Hodyss:2008cg} and CO$_2$--H$_2$O--CH$_3$OH\cite{Ehrenfreund:1999wp} ices in studies with mid-IR spectroscopy.  One of the many advantages of THz-TDS in studying ices, however, is that the features arising from these species in the THz regime result from the collective motion of many molecules (intra-molecular modes) and thus serve as a direct probe of the structure of the ice.\cite{Allodi:2013ma,Ioppolo:2014fd}  This is in contrast to mid-IR spectroscopy, where the ice structure must be indirectly inferred from changes in the lineshape of intermolecular modes.

This utility is immediately obvious from the observations of the linewidths of the $c$-CO$_2$ features within the various mixtures.  While the features from pure $c$-CO$_2$ are relatively sharp, any amount of contamination from CH$_3$OH significantly broadens the transitions.  This demonstrates how the THz transitions of CO$_2$ are useful as a probe of local structure.  As these are solid-phase materials that do not have structural rearrangement happening of a timescale faster than our measurement, the broadening of the observed spectra features results from the different ice environments experienced by the CO$_2$ molecules, and can be correctly characterized as inhomogenously broadened.  Even a 10\% CH$_3$OH contamination is apparently sufficient to create a variety of local environments within the ice, and inhomogeneously broaden these transitions.  

Since the spectral features at THz frequencies are inter-molecular in nature, the amount of inhomogenous broadening offers a direct measure of the number of unique structural environments present in the ice. While the inhomogenous broadening of intra-molecular modes provides insight into the number of different local environments of individual molecules, the inhomogenous broadening of inter-molecular modes can only happen when there are many different structural environments, leading to differences in the frequency of the collective motion of many molecules. 

Interestingly, there is a lack of a clearly increasing trend in linewidth in the THz observations, unlike those in the FTIR.  If we assume that the linewidth from pure $c$-CO$_2$ represents the homogenous value, this suggests that the CO$_2$--CH$_3$OH interaction is somewhat uniform across mixing ratios.  It is possible that the data show a trend of increasing linewidth with increasing CO$_2$ concentration within the ice, but this cannot be claimed definitively given the uncertainties in the measurements. Follow-up studies will examine this effect, as well as providing an in-depth examination of the degree of segregation and the size of the $c$-CO$_2$ domains under various conditions. Finally, the larger degree of broadening observed in the 2.1 THz transition, relatively to the 3.5 THz transition, may provide insight into the nature of these motions in the bulk ice. 

\begin{figure}
\centering
\includegraphics[width=0.5\textwidth]{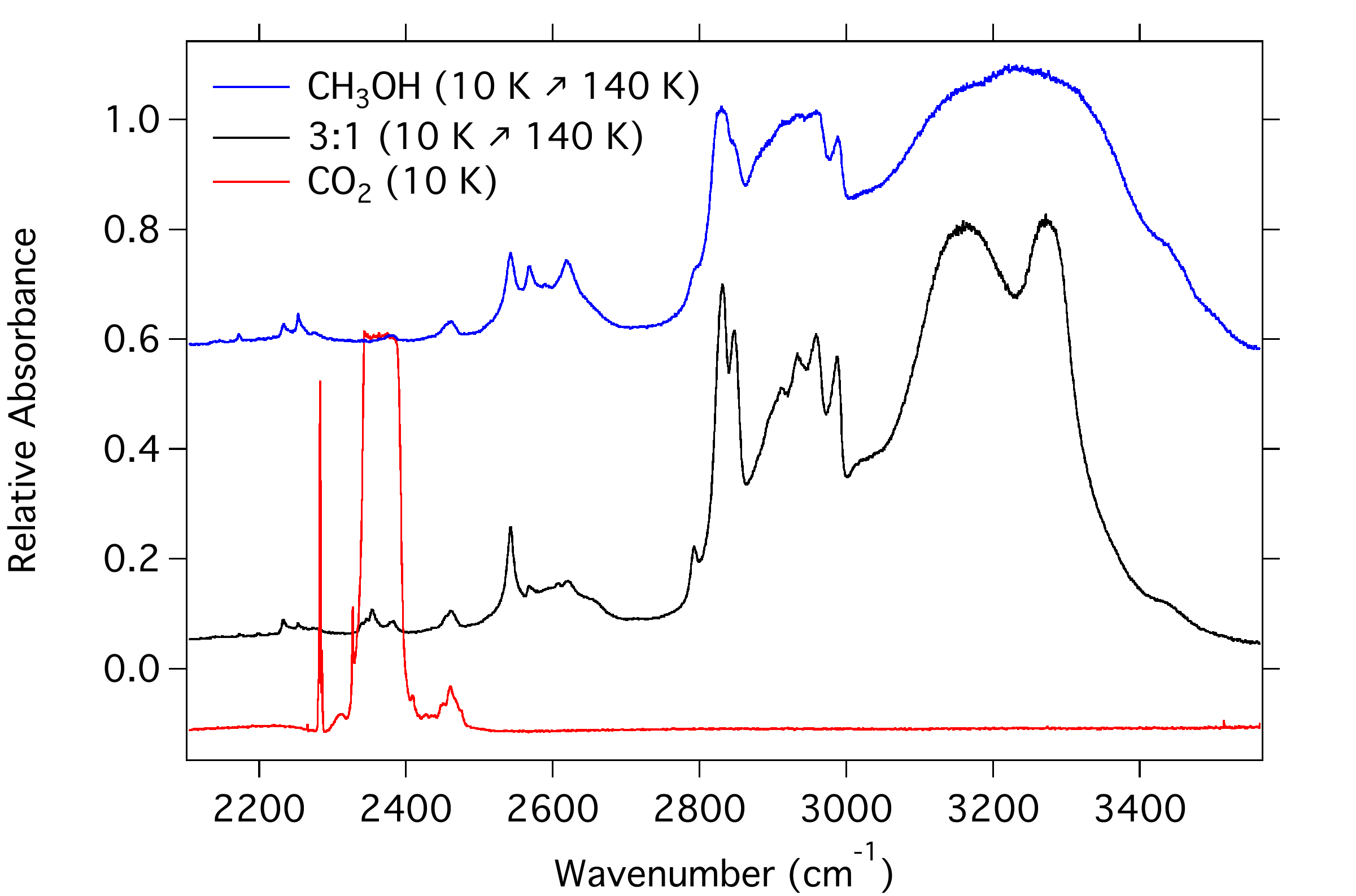}
\caption{FTIR spectra of pure $c$-CO$_2$ at 10 K (bottom), the 3:1 mixture at 10 K after annealing to 140 K (middle), and pure CH$_3$OH at 10 K after annealing to 140 K (top).}
\label{ftir}
\end{figure}

\citet{Hodyss:2008cg} observed that features of $c$-CO$_2$ in their CO$_2$--H$_2$O mixtures initially appear at 60 K, are distinct by 70 K, and gradually lessen as the ice is heated further to 80 K and disappear at 100 K as the CO$_2$ sublimates.  In initially-amorphous CO$_2$--CH$_3$OH ices studied by \citet{Ehrenfreund:1999wp}, CO$_2$ persists in the mixture at temperatures as high as 125 K, and additional spectroscopic features emerge.  These features are likely interactions between CO$_2$ which has crystallized to a degree at lower temperatures, and the CH$_3$OH, which does not begin to crystallize until $\sim$120 K.  This is supported in our data, especially in the 3:1 mixture after annealing to 120 K, where strong $c$-CO$_2$ features remain visible while $c$-CH$_3$OH signal begins to emerge.  Indeed, although the THz signals of $c$-CO$_2$ are obscured by those of $c$-CH$_3$OH, $c$-CO$_2$ is still present in the ice even after annealing to 140 K, as indicated by FTIR spectra recorded contemporaneously (Fig. \ref{ftir}).

Interestingly, three sharp features are observed in the THz spectra of the 3:1 mixture around 2.2--2.8 THz, after the mixture has been annealed to 140 K.  While these features might be present as broad shoulders in the pure CH$_3$OH ice after the same heating process, they are not nearly as distinct.  Perhaps, as was the case with the CO$_2$ stretch at 2340 cm$^{-1}$ observed by \citet{Ehrenfreund:1999wp}, this is an indication of a coupling between the CO$_2$ and CH$_3$OH motions.  

\begin{figure}
\centering
\includegraphics[width=0.5\textwidth]{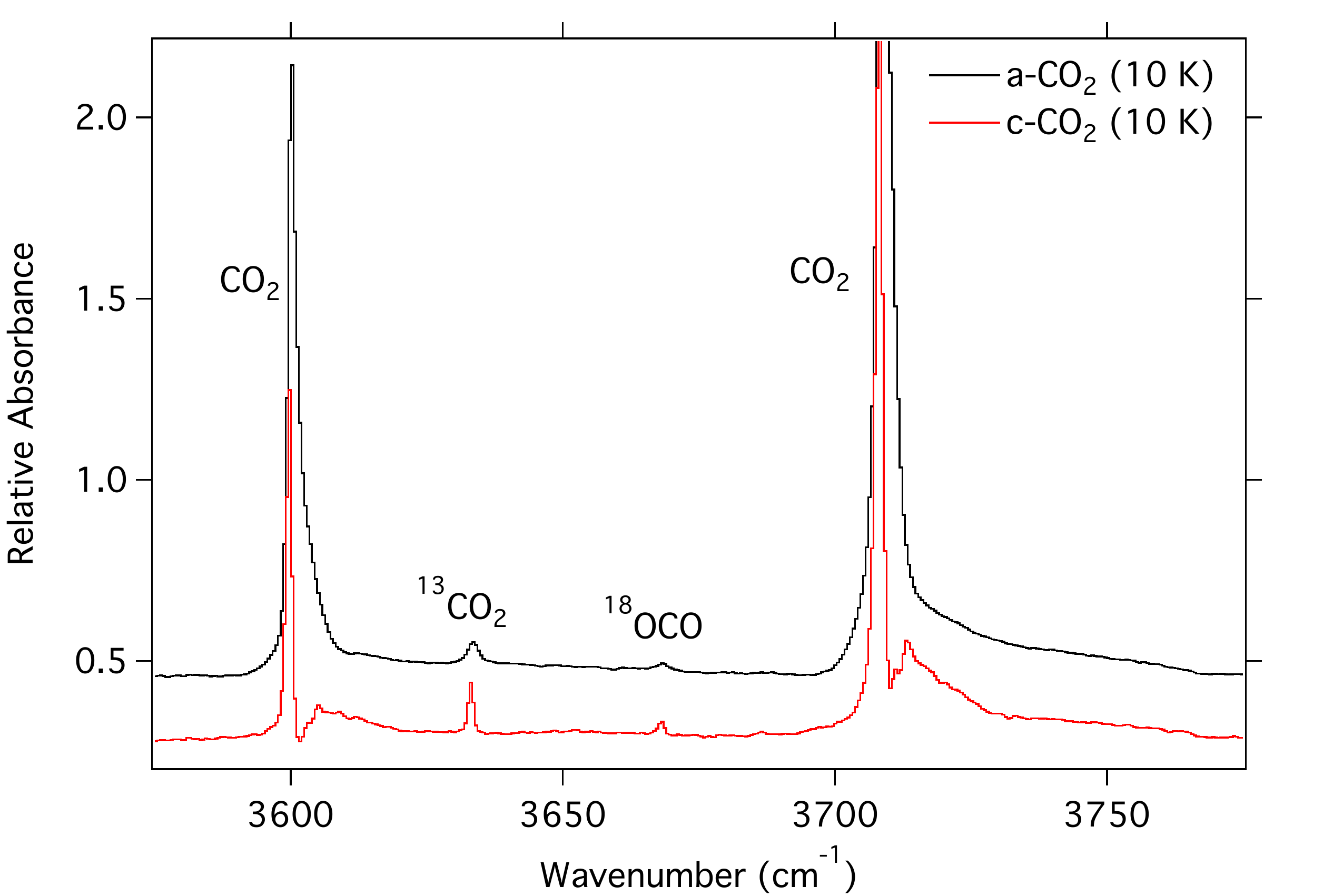}
\includegraphics[width=0.5\textwidth]{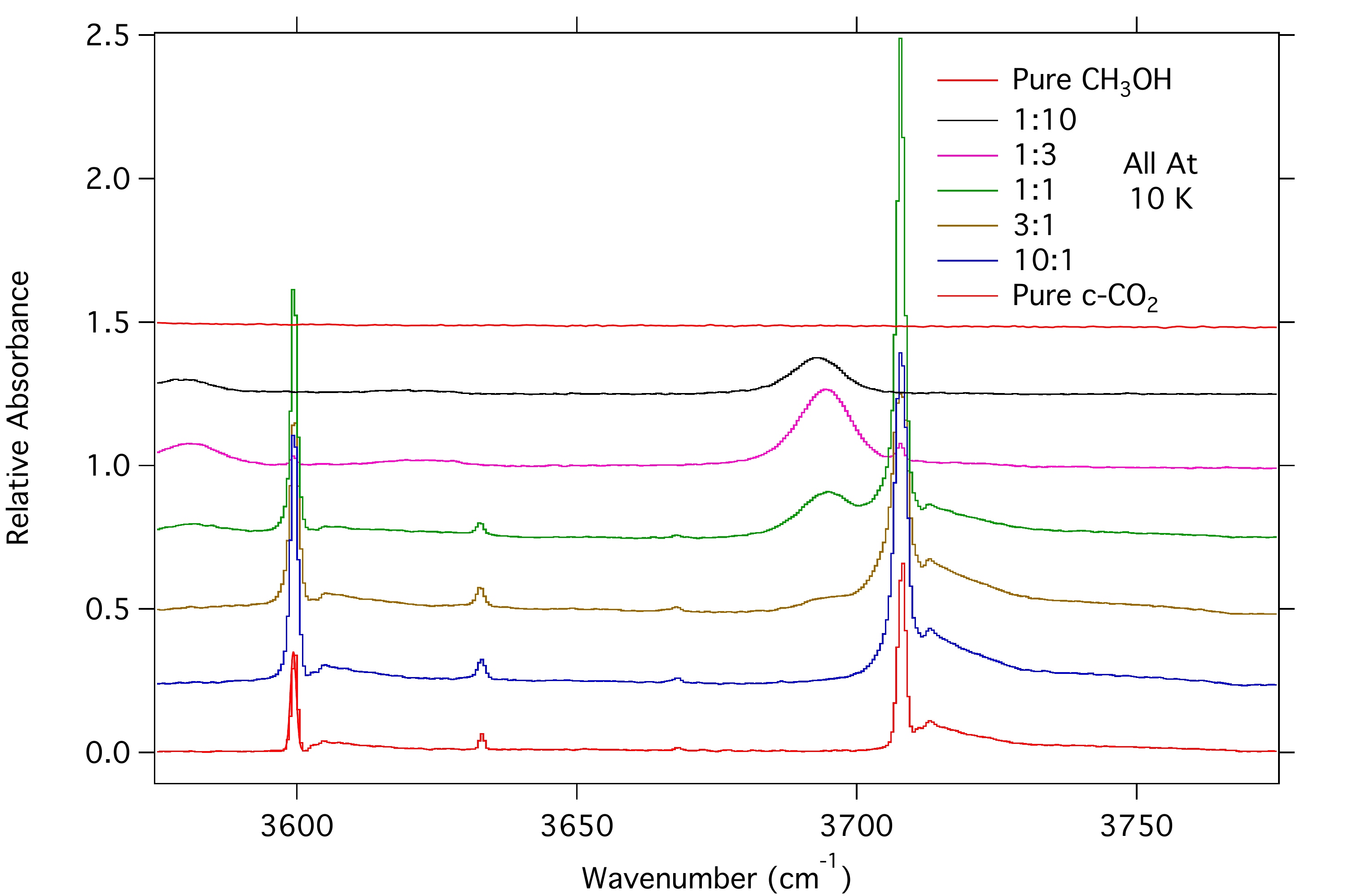}
\includegraphics[width=0.5\textwidth]{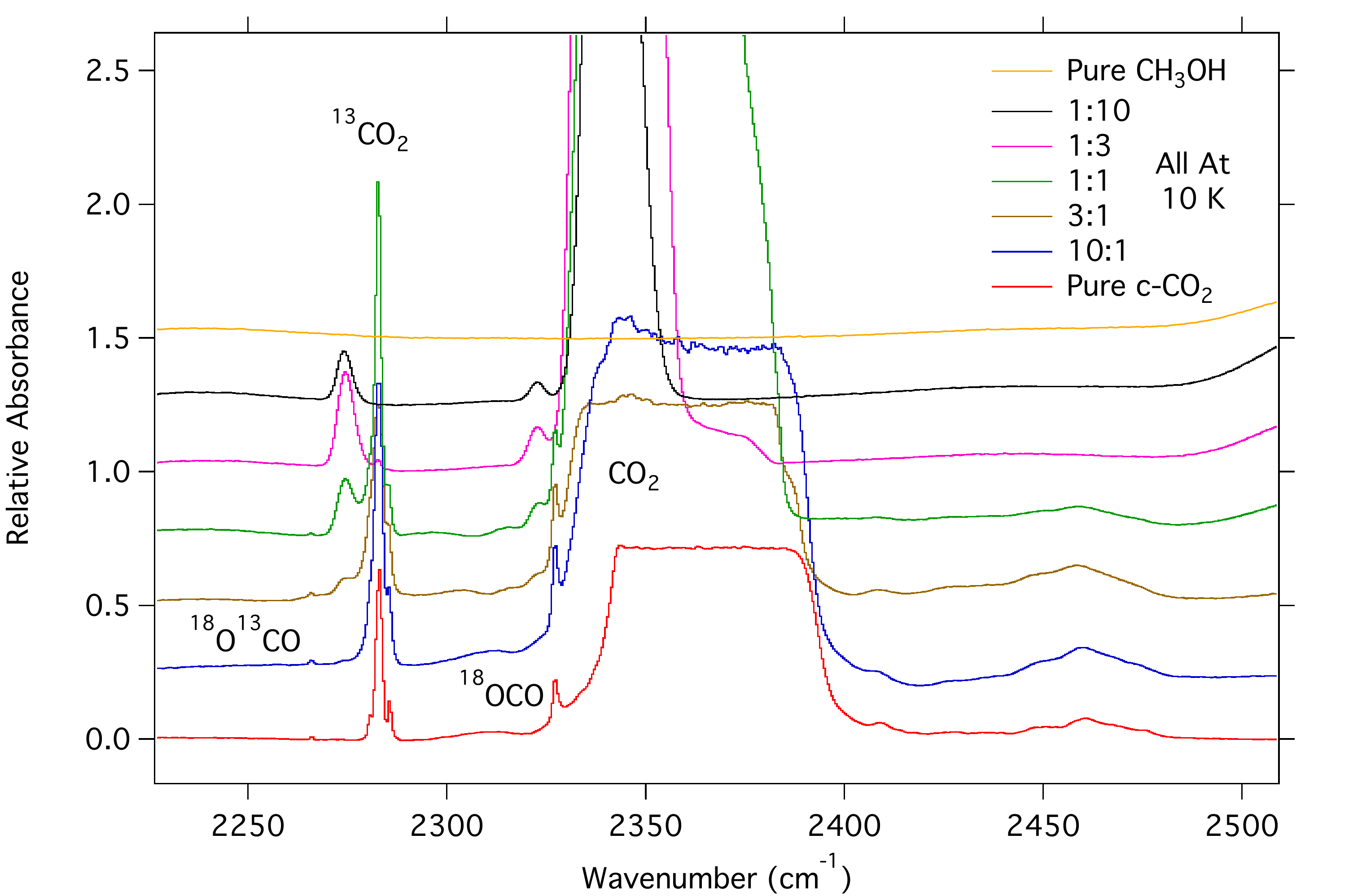}
\caption{(Top) FTIR spectra of the $\nu_1 + \nu_3$ and $2\nu_2 + \nu_3$ combination bands of $a$-CO$_2$ (black) and $c$-CO$_2$ (red) collected in our laboratory at 10 K.  (Middle) The same modes collected during the 10 K experiment (no annealing) for each of the mixtures studied in this work.  (Bottom) The $\nu_3$ band of CO$_2$ and $^{13}$CO$_2$ collected during the 10 K experiment (no annealing) for each of the mixtures studied in this work.  Spectra are vertically-offset for clarity, and several additional features due to isotopologues are indicated.\cite{Lehmann:1977hr,Dartois:2009gw}}
\label{ftirsets}
\end{figure}

In the CH$_3$OH-dominated mixtures, no clear signal from $c$-CO$_2$ is observed in any of the THz spectra, and while the bright $\nu_3$ band is always visible (and saturated) in the FTIR, the combination modes are heavily suppressed in the CH$_3$OH-dominated mixtures (Fig. \ref{ftirsets} middle and bottom).  While the FTIR spectra can distinguish between crystalline and amorphous CO$_2$ under most circumstances (Fig. \ref{ftirsets} top), the THz spectra offer a more powerful probe of segregation within the ice.  The presence of $c$-CO$_2$ features would unambiguously indicate the presence of significantly-sized micro-domains of ordered CO$_2$.  The lack of any such features indicates that little to no aggregation of CO$_2$ domains from within the CH$_3$OH, beyond the initial segregation at deposition, has occurred.  This agrees with the conclusion of \citet{Ehrenfreund:1999wp} that the CH$_3$OH-dominated ices are more thermally-stable, and less subject to reorganization with heating.

Finally, the degree of inhomogeneous broadening of the $c$-CO$_2$, relative to a pure, bulk ice, is an indicator of the variety of size scales of the crystalline micro-domains within the bulk ice.  As the segregated $c$-CO$_2$ approaches a uniform size, and nears the formation of a single bulk crystal, this broadening should reduce to the narrower profiles of the pure $c$-CO$_2$.  A follow-up study could potentially monitor the change in broadening as a function of time while the ice is gently heated and CO$_2$ segregation occurs, offering a direct probe of migration and crystallization timescales and dynamics within the ice.  Further, as the broadening for the 3.5 THz transition relative to the 2.1 THz transition does not appear to be unity, this ratio could be used in astronomical observations to probe the level of segregation, even without a baseline, purely $c$-CO$_2$ signature within the source to set the intrinsic width.

\subsection{Mixture-Dependent Frequencies}

As discussed above, the change in lineshape with mixing ratio in the THz spectra is the most obvious indicator of the effect of the bulk environment on the spectra.  In the FTIR spectra, however, this effect is more readily apparent in the observed frequencies of the transitions.  This is most clear in the case of CO$_2$.  For example, the $\nu_3$ band of CO$_2$ shows a distinct blue-shift as the mixing ratio tends toward pure CO$_2$ (Figure \ref{ftirsets} bottom).  While the $^{12}$CO$_2$ feature is too saturated due to the thickness required for this experiment for a quantitative analysis, the $^{13}$CO$_2$ peak presents a large, 8.7 cm$^{-1}$ shift from the 1:10 to pure $c$-CO$_2$ ices.  An analogous shift has been previously observed for clusters of CO$_2$ in the gas phase.  Here, the increasingly large clusters are essentially a shift toward the purely crystalline CO$_2$ domain, and even larger blue shifts ($\sim$21 cm$^{-1}$) are observed as the cluster grows from the monomer to $N = 13$.\cite{NoroozOliaee:2011hn}

We also note the possibility that signals could arise in the FTIR spectra from combination bands of the long-range, inter-molecular modes in the THz region and the intra-molecular motions in the infrared.  These modes would in theory be distinguishable from simple isotopically-shifted intra-molecular modes by their frequency shifts: such isotopic shifts would be to the red, while combination bands would lie primarily to the blue of the monomer features.  Because our ices are by necessity so thick, and the resulting monomer features saturated, it is likely that the lowest-lying of the combination bands will be buried beneath the overly-wide monomer signal.  Nevertheless, the relatively sharp transitions observed in the THz should have counterparts in the IR, although the frequencies will be shifted and the lineshapes altered due to the effects of vibrational excitation and different coupling to the bulk environment.  Isotopic labeling, which would differentially affect the intra- vs inter-molecular modes and thus make combination bands distinct from monomer features, is a promising avenue to exploring this potential interaction.  Such studies are beyond the scope of the current work, however.

\subsection{Observational Implications}
\label{observational}

In terms of detectability, it seems likely that features of $c$-CO$_2$ will be present in astronomical observations, assuming it is sufficiently segregated both from smaller contaminant species, such as the CH$_3$OH studied here, and from the polar H$_2$O ices where it is formed.  This segregation has already been observed astronomically.\cite{Pontoppidan:2008ew,Kim:2012hz,Escribano:2013gh}  It follows that the THz signals of $c$-CO$_2$ would be excellent targets for interstellar observations, as they are unambiguous evidence of segregated, $c$-CO$_2$.  Given the extensive broadening of the THz modes when the $c$-CO$_2$ is in a mixture, it is also possible that such a width could be used as an indicator of the degree of segregation within these interstellar ices.  Further work will certainly be needed to determine whether this is truly a useful probe, especially at mixing ratios higher than those used in this initial study (i.e. $>10:1$).

\begin{figure}
\centering
\includegraphics[width=0.5\textwidth]{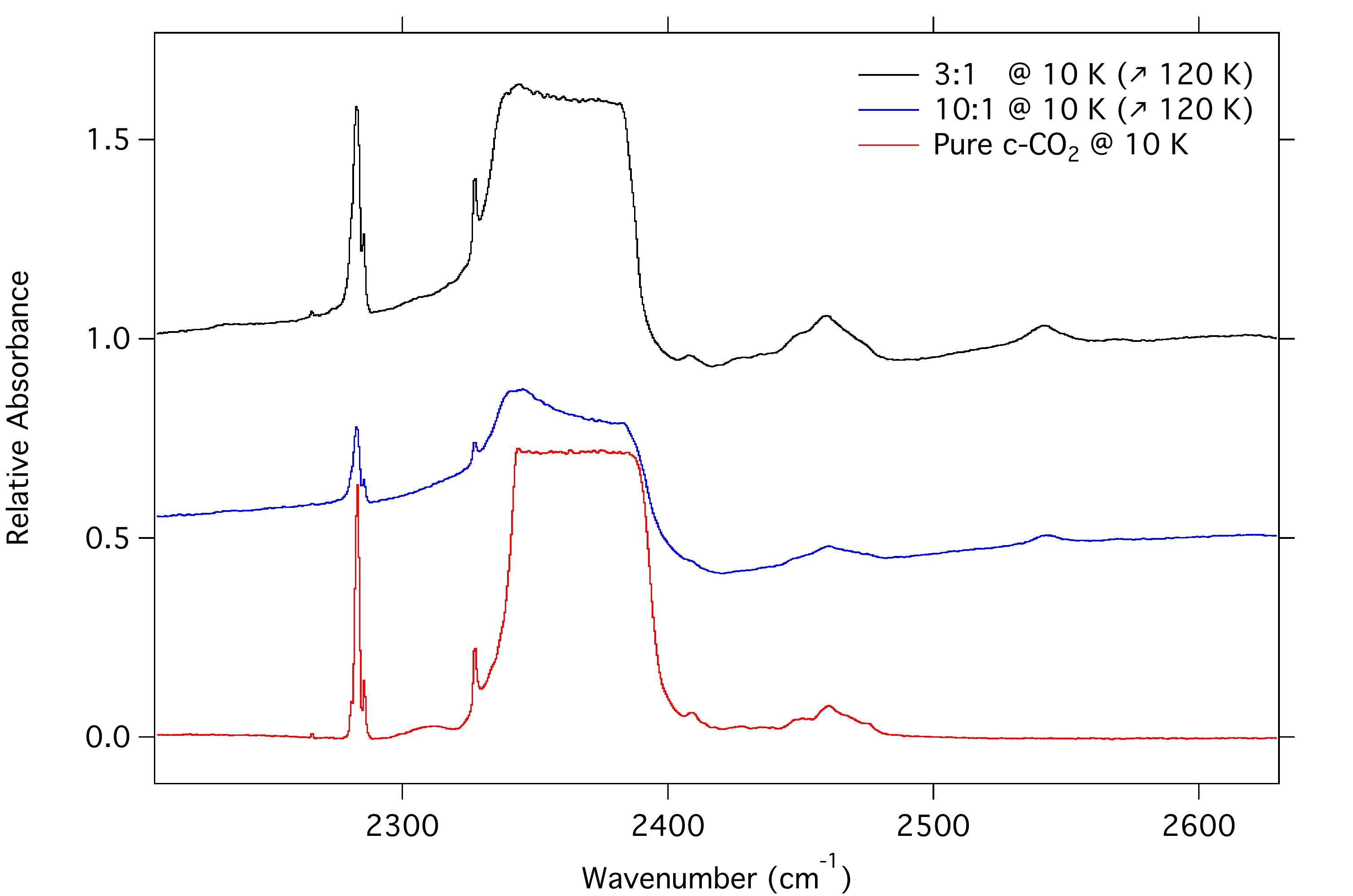}
\caption{FTIR of the 3:1 and 10:1 mixtures after annealing to 120 K showing the remaining CO$_2$ in the ice.  The transitions are saturated in our spectrometer, and the spectra have been vertically-offset for clarity.  A pure $c$-CO$_2$ spectrum at 10 K is provided for reference.}
\label{ftirann}
\end{figure}

Interestingly, the THz features of CH$_3$OH are also distinct after annealing at 120 K, even in both mixtures (10:1 and 3:1) that clearly still contain substantial fractions of CO$_2$ (see Fig. \ref{ftirann}).  Indeed, the $c$-CH$_3$OH features are far stronger than the $c$-CO$_2$, although broader.  Their presence in an interstellar observation would therefore indicate the thermal history of the ices, while the peak positions of the features have already been shown to be dependent on the observed temperature of the ice.\cite{Allodi:2013ma,Ioppolo:2014fd}

An important consideration is the role of dust grain size and composition underlying interstellar ices, and the impact these have on the ice structure and subsequently their spectra.  Indeed, recent modeling work by \citet{Pauly:2016hw} has shown that while the chemical makeup of the ices does not greatly vary with grain size, there can be significant stratification in which size-class of grains is the dominant ice carrier.  For example, during cloud collapse, they find that small grains dominate as ice carriers, resulting in ice thickness of $<$40 ML, versus those of order $\sim$10$^2$ if the model includes only a single grain size.  This raises the important question: to what degree can species segregate within these ices, and how will that affect the spectra they present?

Due to the sensitivity of the second-generation THz spectrometer presented here, we could only study thick laboratory ices (of order $\sim$10$^4$ ML).  The current-generation spectrometer is now capable of studying ices of order $\sim$10$^3$ ML, and the next-generation instrument should push into the 10$^1$ -- 10$^2$ ML regime with the ability to directly measure the optical constants of the ices under investigation.  This is because the electro-optic sampling technique employed here directly measures both the amplitude and phase of the THz pulse.  This presents a experimental approach which is far simpler than complementary techniques with an FTIR that require an asymmetric configuration where the sample resides in an arm of the interferometer.\cite{Birch:1987vj}  Combined with molecular dynamics simulations, radiative transfer, and scattering models over a realistic range in dust particle size, these optical constant measurements will allow for far more accurate modeling of observational spectra.  Such models will in turn provide even more precise information on not just the ice composition and structure, but that of the underlying grain substrate.  With the critical role that the size of icy grains has on the growth of macroscopic grains and gaps in protoplanetary disks,\cite{Zhang:2015id} this information will be essential for understanding physical evolution in star and planet formation.

Indeed, factors such as composition, physical structure and segregation, temperature, and thermal history all play crucial roles in the evolution of molecular complexity both in molecular clouds and in evolving planetary systems.\cite{Garrod:2013id}  Observations of ices in the THz regime offer the potential to shed light on these factors, and in the case of crystallinity and segregation, to do so unambiguously due to the nature of the THz modes.  Molecular ices are the birthplace of the prebiotic complexity which will eventually be incorporated into nascent solar systems, and thus understanding the environments in which they form, and the mechanisms of formation, is critical to understanding the genesis of this primordial material.

\section{Acknowledgements}

The authors thank P. Carroll, I. Finneran, and M. Kelley for their assistance with spectrometer development.  B.A.M. thanks A. Remijan for his support.  The authors thank the anonymous referee for a detailed reading of the manuscript which significantly improved the quality of the work.  S.I. acknowledges funding through a Marie Curie Fellowship (FP1-PEOPLE-2011-IOF-300957) and recent support by the Royal Society.  M.A.A. acknowledges current support from a Yen Postdoctoral Fellowship from the Institute for Biophysical Dynamics at the University of Chicago. The National Radio Astronomy Observatory is a facility of the National Science Foundation operated under cooperative agreement by Associated Universities, Inc.

\footnotesize{
\bibliography{bibliography} 
\bibliographystyle{rsc} 
}

\end{document}